\documentstyle[aps,prb,multicol]{revtex}
\begin{document}
\widetext
\draft

\title{
Wilson's ratio and the spin splitting of magnetic\\
 oscillations in quasi-two-dimensional metals
}

\author{Ross H. McKenzie\cite{email}
}

\address{School of Physics, University of New
South Wales, Sydney 2052, Australia}

\date{\today}
\maketitle
\begin{abstract}
A simple consistency check is proposed for
the Fermi liquid description of the low-temperature
properties of quasi-two-dimensional metals.
In a quasi-two-dimensional Fermi liquid the Zeeman splitting
of magnetic oscillations can be used to determine $g^*$,
the gyro-magnetic ratio which is renormalised  by many-body
effects. It is shown that $g^*/g $  is equal to
Wilson's ratio $R$, the dimensionless ratio
of the low-temperature spin susceptibility to the specific
heat coefficient.                    
Measured values of $g^*/g $ and $R$     are compared
for the layered perovskite Sr$_2$RuO$_4$ and
for a range of organic metals based on the BEDT-TTF molecule.
It is also shown that the 
Pauli paramagnetic limiting magnetic field, at which
singlet superconductivity should
be destroyed by the Zeeman splitting of the electron spins,
is changed from the values given by BCS theory by
a factor of $g^*/g $.
\\
\end{abstract}



\begin{multicols}{2}

Organic molecular crystals based on the BEDT-TTF molecule
are particularly rich low-dimensional electronic systems
in which there is competition between metallic, insulating,
magnetic, and superconducting phases.\cite{wos,ish}
It has recently been argued that the $\kappa$-(BEDT-TTF)$_2$X 
family are strongly correlated electron systems.\cite{fuku,mck}
At temperatures above about 30 K they have a number
of properties that are inconsistent with a Fermi liquid:
the resistivity and thermopower have a non-monotonic
temperature dependence, the mean-free path is less than
a lattice constant, and   there is no Drude peak in 
the optical conductivity.
However, the metallic state at low-temperature has
some properties consistent with a Fermi liquid:
 the resistivity is quadratic in
temperature, the thermopower is linear in temperature,
a Drude peak is present, and  magnetic
oscillations can be observed.\cite{mck}
Similar issues are also relevant
for the layered perovskite Sr$_2$RuO$_4$.\cite{maeno}
It is important to find quantitative measures of the
strength of the correlations and concrete tests
of the extent to which the low-temperature
metallic properties can be described by
a Fermi liquid picture.
In this paper I propose such a test: comparison of
Wilson's ratio to the renormalized g-factor
deduced from the spin splitting of magnetic oscillations.

 Wilson's ratio
has proved to be useful in characterising
strongly correlated Fermi liquids.
In a Fermi liquid
the low-temperature electronic specific heat $C(T)$
is linear in temperature, with a slope $\gamma$.
The magnetic susceptibility $\chi(T)$
is independent of temperature for low temperatures.
Wilson's ratio (also known as the Sommerfeld ratio) is defined
as the dimensionless quantity 
\begin{equation}
 R \equiv {4 \pi^2 k_B^2 \chi(0) \over 3 (g \mu_B)^2 \gamma}
\label{wilson}
\end{equation}
where $g $ is the gyro-magnetic ratio in the absence
of interactions
and $\mu_B$ is the Bohr magneton.
For a non-interacting Fermi gas $R=1$.
In terms of Landau's Fermi liquid parameters,
 $R = (1 + F_0^a)^{-1}$.\cite{vollhardt}
Wilson showed that
for the Kondo model the impurity contributions
to $\chi(0)$ and $\gamma$ give a universal value of $R=2$,
independent of the strength of the interactions.\cite{wilson}
In an isotropic Fermi liquid with a local self energy
(i.e., independent of momentum) it can be shown\cite{bedell}
that $ R<2$,
and for purely repulsive interactions $1< R<2$,
while for purely attractive interactions $R<1$.
In La$_{1-x}$Sr$_x$O$_3$, the doping
induced Mott-Hubbard transition is approached
as $x \to 0$ and
both $\chi(0)$ and $\gamma$ increase significantly with
decreasing $x$ but
$R$ tends to  value of 2, independent of the doping.\cite{tokura}
For heavy fermion metals, even though
$\gamma$ and $\chi(0)$ can be as much as
a hundred times the values predicted by
band structure calculations, $R$ is in the range 0.3 to 2.\cite{rmp}
Although there is some uncertainty in the absolute value
of $R$
(because of uncertainty in the value of the
$g $ that should be used in defining $R$),
the trend is that as one goes from superconducting
to non-magnetic to magnetic heavy fermion materials
$R$ increases.\cite{rmp}
The observed values of $R$ can constrain
theories of strongly correlated Fermi liquids.
For example, certain heavy fermion
models involving strong ferromagnetic correlations
have been ruled out because they predict too large 
a value for $R$.\cite{carlos}
The weak pressure dependence of $R$ for liquid $^3$He has
been used to argue that $^3$He is
``almost localized,'' i.e., close to a Mott-Hubbard
transition, rather than close to a ferromagnetic transition,
as suggested by paramagnon theories.\cite{vollhardt}

I now give a
general expression for $R$ in the case of
a quasi-two-dimensional Fermi liquid with a circular    
Fermi surface of radius $k_F$ within each layer 
and which has a self energy
$\Sigma_\sigma(\omega,k)$ that is independent of the momentum
direction.
$\sigma = \pm 1 $ is the spin index.
The discussion that follows is similar to
Luttinger's treatment of an isotropic three-dimensional
Fermi liquid with a local self energy\cite{lutt}
and Hewson's treatment of the Anderson model.\cite{hewson}
This self energy includes the effect of both
electron-electron interactions {\it and} electron-phonon
interactions.
Let $\epsilon_k = k^2/m_b \simeq k_F (k - k_F)/m_b $
 be the electron dispersion in the absence of interactions where 
$m_b$ is the band mass.
The quasiparticle energies $E_{k\sigma}$ in the presence of
a Zeeman splitting 
$ 2 h= g \mu_B B$
due to a magnetic field $B$ are
given by the roots of
\begin{equation}
E_{k\sigma} - \epsilon_k -{\sigma \over 2}
g \mu_B B - \Sigma_\sigma(E_{k\sigma},k,h)=0
\label{roots}
\end{equation}
We can expand the self energy to first order
in the energy and magnetic field to find that the
quasi-particle energies are
\begin{equation}
E_{k\sigma} = 
{ k_F (k - k_F) \over m^* }     
 -{\sigma \over 2}  g^* \mu_B B 
\label{roots2}
\end{equation}
where $m^*$ is the effective mass, including all many-body effects,
given by
\begin{equation}
  {m_b \over m^*} = {{ 1 - {m_b \over k_F}
 {\partial {\rm Re} \Sigma (0,k_F, 0)\over \partial k}}
\over {1 - {\partial {\rm Re} \Sigma (0,k_F, 0)\over \partial \omega}}}.
\label{zz}
\end{equation}
The renormalised $g$-factor is 
\begin{equation}
{g^* \over g} = 
 {{ 1 - {\partial {\rm Re} \Sigma (0,k_F, 0)\over \partial h}}
\over {1 - {\partial {\rm Re} \Sigma (0,k_F, 0)\over \partial \omega}}}.
\label{gstar}
\end{equation}
The specific heat coefficient and 
spin susceptibility are then       
\begin{equation}
\gamma= {m^* \over m_b} \gamma_0
\ \ \ \ \ \ \ \ \ \ \ \
\chi(0)= {g^* \over g} {m^* \over m_b} \chi_0
\label{susc}
\end{equation}
where $\gamma_0$ and $\chi_0$ are the values in the absence
of interactions.
Hence, Wilson's ratio (\ref{wilson}) is 
\begin{equation}
R= {g^* \over g}
\label{ident}
\end{equation}

I now review how in a quasi-two-dimensional
metal with a single closed Fermi surface
$g^*/g$    can be       determined 
from magnetic oscillations.\cite{gorkov}
The amplitude of
Shubnikov - de Haas and de Haas - van Alphen
oscillations 
is proportional to $R_T R_S$
where $R_S$ is defined below and\cite{sho}
\begin{equation}
R_T = {X \over \sinh X}
\ \ \ \  X = {2 \pi^2 k_B T m^* \over \hbar e B \cos \theta}
\label{split1}
\end{equation}
where $\theta$ is the angle between the magnetic
field and the normal to the conducting layers.
$R_T$ describes how the oscillation amplitude is reduced due to
the phase smearing from finite temperature.
This  temperature dependence 
can be used to determine the
effective mass, $m^*$, 
which can be compared with the band mass $m_b$ calculated
from band structure. In organic metals, the ratio $m^*/m_b$ is often
in the range of 2-5, suggesting large many-body effects.
However, band structure estimates of $m_b$ can differ by as much as
a factor of two\cite{Ching}, and so this does not
represent a very reliable method of determining the
magnitude of many-body effects.
The factor
\begin{equation}
R_S = \cos \left({ \pi S \over 2 } \right) =
 \cos \left( { \pi \over 2 \cos \theta} { g^* m^* \over m_e} \right)
\label{split}
\end{equation}
describes the relative phase of the spin-split Landau levels
where $m_e$ is the mass of a free electron.
$S$ is the ratio of the Zeeman splitting 
 $g^* \mu_B B $ to the cyclotron splitting
 $\hbar \omega_c = \hbar e B \cos \theta /m^*$.
Quasi-two-dimensional systems have the distinct 
advantage that by
 tilting the field the phase factor $S$ can be varied
and $g^* m^*$ and deduced.
This was first done for the two-dimensional electron gas
in a semiconductor heterostructure by Fang and Stiles\cite{stiles}.
It has been done for a range of quasi-two-dimensional
metals\cite{wos} by
finding more than one angle
at which the amplitude of the magnetic oscillations vanishes
(these are referred to as spin-split zeroes).
Table I contains a list of the values of $g^*$ deduced by
this method for various quasi-two-dimensional metals.
The bare g-factor $g$ can be measured by electron spin resonance
and in BEDT-TTF crystals typically has values in the 
range 2.00 to 2.01.\cite{sugano,venturini,nakamura,williams}
Hence, it is sufficient to take $g=2$.

Table II lists the Wilson's ratio $R$ for various quasi-two-dimensional
metals deduced from thermodynamic measurements.
In the $\alpha$ and $\kappa$ 
systems which involve more than one Fermi surface
one might be cautious about comparing these values of $R$ with
the values of $g^*/g$ deduced from magnetic oscillations because
$R$ contains contributions from {\it all} of the Fermi surfaces
whereas $g^*$ is determined for a particular surface.
However, if the self energy is the same on all Fermi surfaces         
then $R$ should equal $g^*/g$.

For five materials values of
both $R$ and $g^*/g$ are available.
For
$\alpha$-(BEDT-TTF)$_2$NH$_4$Hg(SCN)$_4$,      
$\beta$-(BEDT-TTF)$_2$I$_3$,      and
$\beta^{''}$-(BEDT-TTF)$_2$SF$_5$CH$_2$CF$_2$SO$_3$ 
the two values  agree and for 
 $\kappa$-(BEDT-TTF)$_2$Cu(SCN)$_2$
they do not.
For Sr$_2$RuO$_4$ they almost agree.
It is highly desirable that Table II be completed so that
comparisons  can be made for a wide range of materials.
Note that generally $g^*/g$ determined from magnetic
oscillations will have a smaller uncertainty than $R$ 
determined from thermodynamic measurements.
Furthermore, $g^*/g$ can be determined under pressure,
whereas measuring the specific heat coefficient $\gamma$
under pressure
would be extremely difficult.
This provides a useful way of observing how the
correlations vary with pressure.

Finally, I consider how in a Fermi liquid
the Pauli paramagnetic limit is modified
  by many-body effects.
In a spin singlet superconductor for magnetic fields
larger than a critical value $B_P$, the superconductivity
is destroyed by the Zeeman splitting of the electron spins
breaking apart Cooper pairs.\cite{clogston,zuo}
In weak-coupling BCS theory the transition temperature $T_c$
in a field $B$ is given by the solutions of\cite{dupuis}
\begin{equation}
\ln ({ T_c \over T{_{c0}}}) - \Psi ({1 \over 2} ) + {\rm Re} \Psi (
 {1 \over 2}  + { i h \over 2 \pi T_c}
)
=0
\label{pauli}
\end{equation}
where $T_{c0}$ is the transition temperature in zero field 
and $\Psi$ is the digamma function.
This gives  at low temperatures
\begin{equation}
  B_{P}^{BCS} \simeq
{3.6 k_{B}T_{c0} \over g \mu _{B}}.
\label{bpauli}
\end{equation}
How is this result                modified by many-body effects?
In the derivation of (\ref{pauli}) the Matsubura energy
$\epsilon_n= (2n +1 ) \pi T$ is replaced by
 $\epsilon_n  (1 - {\partial {\rm Re} \Sigma (0,k_F,0)\over \partial \omega}) $
and $h$ is replaced by
$ h ( 1 - {\partial {\rm Re} \Sigma (0,k_F,0)\over \partial h})$.  
As a result, I obtain (\ref{bpauli}) with $g$ replaced with $g^*$.
From Table I it can be seen that for most of the
BEDT-TTF superconductors this will be a small correction.
Hence, the many-body effects are not a possible
explaination of the fact that in many organics the upper
critical field, for fields parallel to the layers,
exceeds the paramagnetic limit calculated from BCS theory.\cite{zuo}

In conclusion, it has been shown that the
consistency of a Fermi liquid description of
the low-temperature properties of 
quasi-one-dimensional metals can be
tested by comparing the values of Wilson's ratio
$R$ determined from thermodynamic measurements to values
determined from the spin splitting of magnetic oscillations.
The values obtained may constrain microscopic theories
of the metallic state of these materials.
It should be stressed that the fact that
$R$ is  close to one does not necessarily mean
that the interactions are weak.
This is demonstrated by the fact that there are
heavy fermion materials with values close to one.



This work    was supported by the Australian Research Council.
I  thank J. Wosnitza, J. Schmalian, and J. Merino for helpful discussions.
I am very grateful to H. H. Wang for measuring the susceptibility
of 
$\beta^{''}$-(BEDT-TTF)$_2$SF$_5$CH$_2$CF$_2$SO$_3$
especially for this paper.


\end{multicols}

\newpage

\begin{table}
\caption{
Values of the renormalised $g$-factor for
various quasi-two-dimensional metals, determined from
the spin splitting of magnetic  oscillations.
The effective mass $m^*$ is determined from
the temperature dependence of the 
amplitude of the oscillations (see Eq. (\protect\ref{split1})).
A value of $g=2$ has been used, consistent with electron spin
resonance experiments.
Uncertainties are only given if   they are given in
the original reference.
} \begin{tabular}{llllc} & $m^*/m_e$ & $ g^* m^*/m_e $ & $g^*/g $ & Ref. \\
\tableline
$\alpha$-(BEDT-TTF)$_2$NH$_4$Hg(SCN)$_4$ &
  2.6 $\pm$ 0.1 & 4.45 $\pm$ 0.05  & 0.86 $\pm$ 0.05 
  & \protect\onlinecite{wosnitza2,doporto,sasaki1} \\
$\alpha$-(BEDT-TTF)$_2$KHg(SeCN)$_4$ &
  1.9        & 3.66   & 1.0     & \protect\onlinecite{wosnitza3,sasaki1} \\
$\alpha$-(BEDT-TTF)$_2$TlHg(SeCN)$_4$ &
  2.0 $\pm$ 0.1 & 3.75 $\pm$ 0.02   & 0.94 $\pm$ 0.05
& \protect\onlinecite{goll} \\
$\alpha$-(BEDT-TTF)$_2$KHg(SCN)$_4$  (24 T) &
 1.9\protect\cite{sasaki2} & 3.63                & 1.0  
     & \protect\onlinecite{sasaki1} \\
$\alpha$-(BEDT-TTF)$_2$KHg(SCN)$_4$ (below 23 T) &
 1.6 \protect\cite{sasaki2} & 4.7      & 1.5              
     & \protect\onlinecite{sasaki1} \\
 $\kappa$-(BEDT-TTF)$_2$Cu(SCN)$_2$
 &  3.3 $\pm$ 0.3    &   5.2  &  0.8   & \protect\onlinecite{wosnitza2} \\
 $\kappa$-(BEDT-TTF)$_2$Cu[N(CN)$_2$]Br
 &  5.4 $\pm$ 0.1 \protect\cite{mielke},
 6.4 $\pm$ 0.5 \protect\cite{weiss}&     &     & \\
$\kappa$-(BEDT-TTF)$_2$I$_3$ 
 &  3.9 $\pm$ 0.1    &  8.63  &  1.14 
 & \protect\onlinecite{schweitzer} \\
$\kappa$-(BETS)$_2$GaCl$_4$ 
 &  3.27             &  6.6   &  1.0 
 & \protect\onlinecite{pesotskii} \\
$\beta_H$-(BEDT-TTF)$_2$I$_3$ 
 &  4.2 $\pm$ 0.2    &  12.0  &  1.5   & \protect\onlinecite{wosnitza7} \\
$\beta$-(BEDT-TTF)$_2$IBr$_2$ 
 &  4.0     &  9.0  &  1.15   & \protect\onlinecite{wosnitza6} \\
$\beta^{''}$-(BEDT-TTF)$_2$SF$_5$CH$_2$CF$_2$SO$_3$
&  1.90 $\pm$ 0.05       & 3.90 $\pm$ 0.05   & 1.00 $\pm$ 0.03  
  & \protect\onlinecite{wosnitza5} \\
Sr$_2$RuO$_4$ 
&  4.3                   & 10.6          & 1.2 
  & \protect\onlinecite{yoshida} \\
\end{tabular}
\label{table1}
\end{table}

\begin{table}
\caption{
Wilson's ratio $R$ for various quasi-two-dimensional
 metals deduced from measurements
of the specific heat coefficient $\gamma$ and
the spin susceptibility $\chi(0)$.
These values of $R$ can be compared to the values of $g^*/g$ given
in Table I.
(In the units used here, Eq. (\protect\ref{wilson}) with $g=2$
becomes $ R = 0.73 \chi(0)/\gamma$).}
 \begin{tabular}{llll}
 & $\gamma$ (mJ/(K$^2$ mol))& $ \chi(0)$ (10$^{-5}$emu/mol) & $R $  \\
\tableline
$\alpha$-(BEDT-TTF)$_2$NH$_4$Hg(SCN)$_4$ &
26 $\pm$ 1 \protect\cite{nakazawa},
29 $\pm$ 2 \protect\cite{andraka3}
 & 
24 $\pm$ 2 \protect\cite{miyagawa}
& 0.7 $\pm$ 0.2 \\
 $\kappa$-(BEDT-TTF)$_2$Cu(SCN)$_2$  &
25 $\pm$ 3 \protect\cite{andraka2}
& 43 $\pm$ 3  \protect\cite{nakamura,haddon} & 
1.4 $\pm$ 0.2 \\
 $\kappa$-(BEDT-TTF)$_2$Cu[N(CN)$_2$]Br &
 22 $\pm$ 3 \protect\cite{andraka} &  43 $\pm$ 3
\protect\cite{haddon} & 
1.4 $\pm$ 0.2 \\
 $\kappa$-(BEDT-TTF)$_2$I$_3$  &
19 $\pm$ 1.5 \protect\cite{wosnitzax}
& & \\
$\beta$-(BEDT-TTF)$_2$I$_3$  &
24 $\pm$ 3  \protect\cite{stewart}
& 44 $\pm$ 3  \protect\cite{venturini,haddon,roth} &
1.4 $\pm$ 0.2  \\
$\beta$-(BEDT-TTF)$_2$IBr$_2$  &
& 75 $\pm$ 5  \protect\cite{venturini2}& \\
$\beta^{''}$-(BEDT-TTF)$_2$SF$_5$CH$_2$CF$_2$SO$_3$ &
19 $\pm$ 1 \protect\cite{wanka} &
26 $\pm$ 4 \protect\cite{wang}  & 1.0 $\pm$ 0.2\\
Sr$_2$RuO$_4$  &
37.4       \protect\cite{mack} &
76  \protect\cite{ishida}  & 1.5 \\
\end{tabular}
\label{table2}
\end{table}


\begin{references}

\bibitem[*]{email} email: ross@phys.unsw.edu.au

\bibitem{wos}
J. Wosnitza,
{\it Fermi Surfaces of Low Dimensional
 Organic Metals and Superconductors}
(Springer, Berlin, 1996).

\bibitem{ish} T. Ishiguro, K. Yamaji,
and G. Saito,
{\it Organic Superconductors}, Second Edition
(Springer, Berlin, 1998).

\bibitem{fuku}
 H. Kino and H. Fukuyama, J. Phys. Soc. Jpn.
{\bf 65}, 2158 (1996);
K. Kanoda,
 Physica C {\bf 282}, 299 (1997).

\bibitem{mck}
R. H. McKenzie, Science {\bf 278}, 820 (1997);
Comments Cond. Mat. Phys.
{\bf 18}, 309 (1998).

\bibitem{maeno}
Y. Maeno {\it et al.},
J. Phys.   Soc.  Jap. {\bf 66}, 1405 (1997);
A. W. Tyler, A. P. Mackenzie, S. Nishizaki, and
Y. Maeno, Phys. Rev.  B {\bf 58}, R10107 (1998).

\bibitem{vollhardt}
D. Vollhardt,  
Rev. Mod. Phys. {\bf 56}, 99   (1984).

\bibitem{wilson}
K. G. Wilson,
Rev. Mod. Phys. {\bf 47}, 773  (1975).

\bibitem{bedell}
J. R. Engelbrecht and K. S. Bedell,
Phys. Rev. Lett. {\bf 74}, 4265  (1995).

\bibitem{tokura}
Y. Tokura  {\it et al.},
Phys. Rev. Lett. {\bf 70}, 2126  (1993).

\bibitem{rmp}
G. R. Stewart,
Rev. Mod. Phys. {\bf 56}, 755  (1984).

\bibitem{carlos}   C.  Sanchez-Castro and K. S. Bedell,
 Phys. Rev. B {\bf 43}, 12874 (1991),
and references therein.

\bibitem{lutt}
J. M. Luttinger,              
Phys. Rev.  {\bf 119}, 1153  (1960).

\bibitem{hewson}
A. C. Hewson,
{\it The Kondo Problem to Heavy Fermions,}
(Cambridge, Cambridge, 1993), p. 116.

\bibitem{gorkov}
A similar discussion for isotropic three-dimensional
Fermi liquids was given by
Y. A. Bychkov  and L. P. Gorkov,
Sov. Phys. JETP {\bf 14}, 1132 (1962).

\bibitem{sho}
D. Shoenberg, {\it Magnetic Oscillations in
 Metals}, (Cambridge, Cambridge, 1984).

\bibitem{Ching}Y.-N. Xu  {\it et al.},
Phys. Rev. B {\bf 52}, 12946 (1995); {\bf 55}, 2780 (1997).

\bibitem{stiles}
F. F. Fang and P. J. Stiles, 
Phys. Rev.  {\bf 174}, 823   (1968).

\bibitem{sugano}T. Sugano, G. Saito, and M. Kinoshita,     
Phys. Rev. B {\bf 35}, 6554  (1987).

\bibitem{venturini}
E. L. Venturini {\it et al.},
Phys. Rev. B {\bf 32}, 2819  (1985).

\bibitem{nakamura} T. Nakamura {\it et al.},
J. Phys. Soc. Japan {\bf 63},  4110  (1994).

\bibitem{williams}
J. M. Williams {\it et al.},
{\it Organic superconductors (including fullerenes) : synthesis,
structure, properties, and theory}
(Prentice Hall, Englewood Cliffs, 1992), p. 184.

\bibitem{clogston}
A.M. Clogston, Phys. Rev. Lett. {\bf 9}, 266 (1962);
 B.S. Shandrasekhar, Appl. Phys. Lett. {\bf 1}, 7 (1962).

\bibitem{zuo}
For a comprehensive discussion of paramagnetic limiting
in organic superconductors see F. Zuo {\it et al.}, cond-mat/9904186,
and references therein.

\bibitem{dupuis}   N.  Dupuis, Phys. Rev. B {\bf 51}, 9074 (1995).

\bibitem{wosnitza2}
J. Wosnitza {\it et al.},
Phys. Rev. B {\bf 45}, 3018 (1992).

\bibitem{doporto}
M. Doporto  {\it et al.},
Phys. Rev. Lett. {\bf 69}, 991  (1992).

\bibitem{sasaki1}
T. Sasaki and T. Fukase,
Phys. Rev. B {\bf 59}, 13874 (1999).

\bibitem{wosnitza3}
J. Wosnitza {\it et al.},
Synth. Metals {\bf 85}, 1479 (1997).

\bibitem{goll}
G. Goll, 
J. Wosnitza, and N. D. Kushch,
Europhys. Lett. {\bf 35}, 37   (1996).

\bibitem{sasaki2}
T. Sasaki, W. Biberacher, and T. Fukase,
Physica    B {\bf 246-247}, 303   (1998).

\bibitem{mielke} C. H. Mielke      {\it et al.},
Phys. Rev. B {\bf 56},  R4309 (1997).

\bibitem{weiss}  H. Weiss       {\it et al.},
JETP Lett.   {\bf 66},  202   (1997).

\bibitem{schweitzer}
M. Heinecke, K. Winzer, and D. Schweitzer,
Z. Phys. B       {\bf 93}, 45  (1993);
E. Balthes {\it et al.},
Z. Phys. B       {\bf 99}, 163 (1996).

\bibitem{pesotskii}
S. I. Pesotskii {\it et al.},
JETP         {\bf 88}, 114  (1999).

\bibitem{wosnitza7}
D. Beckmann {\it et al.},
Z. Phys. B {\bf 104}, 207 (1997).

\bibitem{wosnitza6}
J. Wosnitza {\it et al.},
J. Phys. I France {\bf 6,} 1597 (1996).

\bibitem{wosnitza5}
D. Beckmann {\it et al.},
Eur. Phys. J. B {\bf 1}, 295 (1998).

\bibitem{yoshida} Y. Yoshida  {\it et al.},
J. Phys. Soc. Japan {\bf 67},  1677  (1998).

\bibitem{nakazawa}Y. Nakazawa, A. Kawamoto, and K. Kanoda,
Phys. Rev. B {\bf 52}, 12890 (1995).

\bibitem{andraka3}
B. Andraka {\it et al.},
Phys. Rev. B {\bf 42}, 9963  (1990).

\bibitem{miyagawa}K. Miyagawa, A. Kawamoto, and K. Kanoda,
Synth. Metals {\bf 86}, 1987 (1997).

\bibitem{andraka2}
B. Andraka {\it et al.},
Phys. Rev. B {\bf 40}, 11345 (1989).

\bibitem{haddon}
R. C. Haddon, A. P. Ramirez, and S. H. Glarum,
Adv. Mater. {\bf 6}, 316 (1994).





\bibitem{andraka}
B. Andraka {\it et al.},
Solid State Commun. {\bf 79}, 57 (1991).




\bibitem{wosnitzax}
J. Wosnitza {\it et al.},
Phys. Rev. B {\bf 50}, 12747 (1996).

\bibitem{stewart}
G. R. Stewart {\it et al.},
Phys. Rev. B {\bf 33}, 2046 (1986).

\bibitem{roth}B. Rothaemel {\it et al.},
Phys. Rev. B {\bf 34}, 704 (1986).

\bibitem{venturini2}
E. L. Venturini {\it et al.},
Synth. Metals {\bf 27}, A243 (1988).

\bibitem{wanka}
S. Wanka  {\it et al.},
Phys. Rev. B {\bf 57}, 3084 (1998).

\bibitem{wang}
H. H. Wang, unpublished.

\bibitem{mack} A. P. MacKenzie  {\it et al.},
J. Phys. Soc. Japan {\bf 67},  385   (1998).

\bibitem{ishida}
K. Ishida {\it et al.},
Phys. Rev. B {\bf 56}, R505 (1997).



\end{references}
\end{document}